\documentclass[twocolumn,showpacs,preprintnumbers,amsmath,amssymb]{revtex4}

\usepackage{graphicx}
\usepackage{dcolumn}
\usepackage{bm}

\begin{document}
\bibliographystyle{prsty}


\title{Anomalous drift of spiral waves in heterogeneous excitable media}

\author{S. Sridhar$^1$, Sitabhra Sinha$^1$ and Alexander V. Panfilov$^2$}
\affiliation{%
$^1$The Institute of Mathematical Sciences, CIT Campus, Taramani,
 Chennai 600113, India.}
\affiliation{%
$^2$Department of Theoretical Biology, Utrecht University, Padualaan 8, Utrecht 3584 CH, The Netherlands
}%

\date{\today}

\begin{abstract}
We study the drift of spiral waves in a simple model of heterogeneous
excitable medium,
having gradients
in local excitability or cellular coupling.
For the first time, we report the anomalous drift of spiral waves towards
regions having higher excitability, in contrast to all earlier
observations in reaction-diffusion
models of excitable media.
Such anomalous drift can promote the onset of 
complex spatio-temporal patterns, e.g., those
responsible for life-threatening arrhythmias in the heart.
\end{abstract}

\maketitle

Spatial patterns of activity, in particular spiral waves,
are observed 
in a broad class of physical, biological and chemical excitable 
systems~\cite{Keener98}. One of the most important contexts in which
spiral waves occur is that of electrical activity in the heart,
where they can  act as local sources of high-frequency excitations. This
disrupts the rhythmic pumping action of the heart,
leading to irregularities
known as arrhythmias~\cite{Davidenko91}. Understanding the dynamics of 
spiral waves may potentially result in improved methods for controlling such
arrhythmias~\cite{Gray95,Barkley92,PanfH:93,PanfKen:93c,Sinha01,Shajahan07}.
Spiral wave dynamics, primarily characterised by the motion of its core (i.e.,
the trajectory of the spiral wave tip, defined to be a phase singularity)
can be either stationary rotation, or, evolving with time as in the
case of meandering and drift~\cite{Garfinkel2000}.
Drift, which has 
a significant linear translational component,
is a possible underlying mechanism for
polymorphic ventricular tachycardia~\cite{Gray95,Garfinkel2000}.
This arrhythmia, which is characterised by an aperiodic electrocardiogram,
can be a precursor of fully
disordered activity that characterizes potentially fatal
ventricular fibrillation~{\cite{Oxford05}.
Therefore, understanding the mechanisms leading to spiral wave drift 
is not only a problem of central interest for physics of excitable media,
but also has potential clinical significance.

One of the most important causes of spiral drift is the heterogeneous
nature of the 
excitable medium. This
was first predicted
in cellular automata models with step-like or discontinuous 
inhomogeneity~\cite{Krinsky68}, 
which was later
confirmed by experiments~\cite{Pertsov90,Markus92}.
Subsequently, drift has been shown to be induced by a smooth gradient of 
excitability in both simple and biologically 
realistic ionic models of cardiac tissue~\cite{Panfilov83,Panfilov91,Panfilov03}.
Theoretical arguments indicate that the direction of the transverse
component of the spiral drift (i.e., orthogonal to the gradient) 
depends on model 
parameters~\cite{Ivanitsky89}. On the other hand, the longitudinal component 
is always directed towards the region with lower excitability (corresponding
to longer periods of spiral rotation)~\cite{Panfilov03}. 
This phenomenon has been seen in a variety of excitable
media models 
of different complexity ~\cite{Panfilov83,Panfilov91,Panfilov03}.
However, till date 
there is no satisfactory understanding of the reasons behind the
spiral wave drift towards regions with longer rotation period.
Although earlier kinematic studies suggested the possibility of
drift towards region with higher excitability~\cite{Mikhailov94},
it has never actually been observed in a model of excitable
tissue.
The occurrence of drift in the direction of shorter period 
(corresponding to higher excitability) may have clinical significance,
as it moves
the spiral wave to a section where it rotates faster.
This increases the likelihood of onset of additional wavebreaks
away from the core, in the region where the medium is less excitable.
Thus, it is a possible generation mechanism for ``mother rotor" 
fibrillation~\cite{Gray96,Jalife02,Panfilov09},
characterised by a stationary persistent source of high-frequency
excitations giving rise to turbulent activity in the heart.

Electro-physiological heterogeneities in cardiac tissue 
may arise, in general, through spatial variation in the 
ionic
currents 
of the excitable  cells.
There can also be gradients in the inter-cellular coupling
as a result of the
inhomogeneous distribution of the conductances of gap junctions
connecting neighboring cells. 
In this paper, we use a simple model of cardiac tissue to investigate 
the role of both these types
of heterogeneities in governing the direction of 
the spiral wave drift.
We report the existence of a regime where the spiral wave core moves towards
the region with higher excitability.
This is a novel 
finding,  never before observed in a model of excitable
media and 
it may substantially increase the understanding of how heterogeneities
affect spiral wave dynamics in the heart.

A generic model of excitable media that describes the dynamics
of trans-membrane potential $V$ in cardiac tissue has the form
\begin{equation} \label{eq1}
\partial V/ \partial t    = {\nabla} \gamma D{\nabla}V + \alpha I_{ion}(V,g_i),
\end{equation}
\begin{equation}
\partial g_i/ \partial t    = F(V,g_i).
\label{eq2}
\end{equation}
Here, $I_{ion}$ is the total ionic current traveling through the channels
on the cellular membrane, $D$ accounts for the inter-cellular coupling 
and $g_i$ describes the
dynamics of gating variables for the various ion channels. 
In this paper, we study the effects of heterogeneous distribution of the 
ionic currents and intercellular couplings. For this purpose, we introduce 
the parameters
$\alpha$ and $\gamma$, which represent the spatial variation
in ionic currents and conduction properties (respectively) for an 
inhomogeneous medium. Parameter $\alpha$ directly scales the value of 
the ionic current in Eq.~\ref{eq1}, while $\gamma$ scales 
the diffusion coefficient as $D=D_0+\gamma(x)$
($D_0=1$ for the rest of the paper).
In this study, we have used the Barkley model~\cite{Barkley90}, where the
several gating variables are aggregated into a single variable $g$
that controls the slow recovery dynamics of the medium with $F(V,g) = V-g$. 
The
nonlinear dependence of the ionic current on the fast variable $V$
is represented by the cubic function 
$I_{ion} = [V(1-V)(V-((g+b)/a)]/ \epsilon$, where $a$ and $b$ are parameters
governing the local kinetics and $\epsilon$ is the relative time scale between
the local dynamics of $V$ and $g$.
The spatial heterogeneity of local excitability and cellular coupling are
assumed to have linear functional form, viz., $\alpha (x) = \alpha_0 + 
\Delta\alpha~x$ and $\gamma (x) = \gamma_0 + \Delta\gamma~x$. The
variable $x$ represents the spatial position along the principal direction
of the inhomogeneity gradient, the origin being considered to be the
initial position of the spiral wave tip. At this point, $\alpha = \alpha_0$, 
$\gamma = \gamma_0$, and $\Delta\alpha, \Delta\gamma$ measure their rate of
change along the gradient. For all the figures in this paper, we have used
$\alpha_0=1.15$ and $\gamma_0=1.3$. 

The two dimensional system is discretized on a square spatial grid of
size $L \times L$ ($L = 200$ for the figures shown here). 
The values of space step 
$\Delta x$ and time step $\Delta t$ used are $0.5$ and $0.005$ respectively. 
A sample of simulations have been  repeated for $\Delta x = 0.25$ to verify 
numerical
accuracy. The model equations are solved using forward Euler scheme with a 
standard 5-point stencil for the Laplacian. No-flux boundary conditions 
are implemented
at the edges of the simulation domain. 
The initial condition for all simulations is
a stable non-meandering spiral, the spiral tip being at 
the centre of the simulation
domain. 

Increasing either the ionic current (via $\alpha$) or inter-cellular
coupling (via $\gamma$) 
results 
in increasing the excitability of the medium.
Thus, to investigate the role of heterogeneity in spiral drift, we
have considered spatial gradients in $\alpha$ or $\gamma$ individually
(keeping the other parameter constant).
After extensive numerical simulations that scan over the ($a,b$) 
parameter space of the Barkley model, we have found that it is 
indeed possible to observe
{\em anomalous drift} of the spiral, i.e., drift towards regions with 
higher excitability (shorter periods).
An example of such anomalous drift is shown in Fig.~\ref{Figure1}~(A,C). 
For comparison, in Fig.~\ref{Figure1}~(B,D) we show the normal drift 
of the spiral towards regions of lower excitability. This is seen for
a set of $(a,b)$ values which is farther from the boundary with the 
sub-excitable region of the Barkley model~\cite{Pumir09} than the 
$(a , b)$ parameter 
set for which anomalous drift is observed in Fig.~\ref{Figure1}~(A,C).

\begin{figure}
\centerline{
\includegraphics[width=1.0\linewidth,clip]{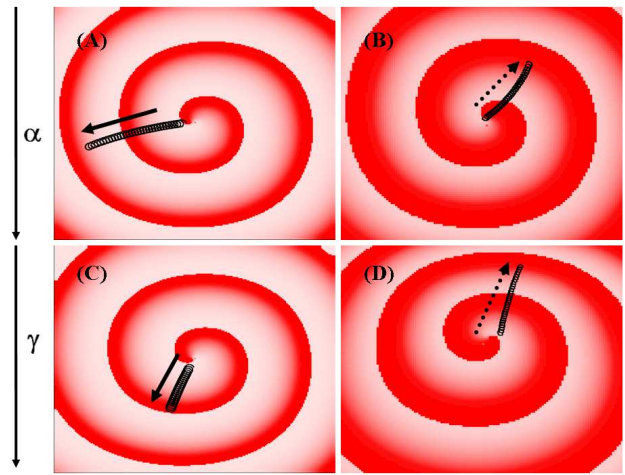}
}
\caption[]{{\bf Anomalous and normal drift of spiral wave.
} 
Pseudocolor images of spiral wave at the instant when the gradient in
local excitability $\alpha$ (top row) and cellular coupling (bottom row)
is applied. 
(A, C) Anomalous drift in the direction of higher excitability, i.e., 
towards increasing values of $\alpha$ (A) or $\gamma$ (C), 
the direction being shown by solid arrows. 
(B, D) Normal drift in  $\alpha$ (B) or $\gamma$ (D) gradient.
Parameter values are $a = 0.82$, $b = 0.13$ (for A, C)
and  $a = 1.02$, $b = 0.15$ (for B, D).
The gradients applied are (A, B) $\Delta \alpha = 0.0025$, 
$\Delta \gamma = 0$, and (C, D)  $\Delta \alpha = 0$, $\Delta \gamma 
= 0.020$. In all cases, the gradient is along the vertical direction, 
with $\alpha$ or $\gamma$ increasing from top to bottom.
In (B,D) the region around the core is magnified to make the wavelength
of the spiral comparable to that in (A,C).
}
\label{Figure1}
\vspace{-0.7cm}
\end{figure}

To analyse the genesis of anomalous drift, we first look at
how the parameters $\gamma$ and $\alpha$ affect 
the spiral wave in an {\em homogeneous} medium.
As $\gamma$ is only a scaling factor for the diffusion coefficient, 
the period of the spiral wave does not depend on it. Further, scaling
arguments suggest that the spiral wavelength increases as a square root 
of $\gamma$. 
Thus, for normal drift in the presence of a gradient in $\gamma$, the
spiral moves towards the shorter wavelength region, while 
for anomalous drift, it is directed towards longer wavelengths.
Fig.~\ref{Figure2} shows the variation of the spiral period and  wavelength 
as a function of the parameter $\alpha$, both of which decrease as
$\alpha$ increases~\cite{note1}. From these results
we can infer that, for normal drift in the presence of $\alpha$
gradient, the period and wavelength of the spiral
increase as the core moves towards lower $\alpha$ regions. In contrast, 
we see a decrease in the period and wavelength in the case of anomalous drift 
towards regions having higher values of $\alpha$.

\begin{figure}
\centerline{
\includegraphics[width=8cm]{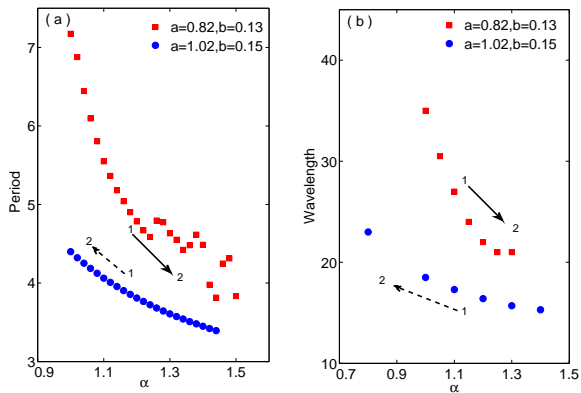}
}
\caption[]{{\bf The variation of spiral period and wavelength 
as a function of the parameter $\alpha$.
}
The symbols ``1" and ``2" 
correspond to the values of $\alpha$
in the region around
the initial and final positions (respectively) of the spiral waves in 
Fig.~\ref{Figure1}, with the same sets of Barkley model parameters being
used.
The solid and broken arrows represent the directions
of anomalous and normal drift, respectively, in presence of a gradient
in $\alpha$.
}
\label{Figure2}
\vspace{-0.3cm}
\end{figure}

Next, we study the effect of the magnitude of spatial gradient in
$\alpha$ or $\gamma$
on the velocity of spiral drift.
Fig.~\ref{Figure3} shows the longitudinal component of the drift
velocity, $V_L$, i.e., along the gradient, 
as a function of the spatial variation in 
$\alpha$ or $\gamma$. Note that, positive $V_L$ corresponds to
anomalous, while, negative $V_L$ corresponds to normal drift of 
the spiral wave.
Fig.~\ref{Figure3} shows that, for normal drift, increasing either of
the
gradients results in a monotonic increase of $V_L$. 
However, 
in the case of anomalous drift as a result of $\alpha$ gradient, 
we see  a {\em non-monotonic} behavior in $V_L$,
which first increases but then  decreases and becomes  negative 
(Fig.~\ref{Figure3},~a).
Thus, the anomalous
drift of the spiral towards higher excitability in  $\alpha$ gradient is
seen only for small $\Delta\alpha$. For higher $\Delta\alpha$, there
is a reversal of direction and the spiral exhibits normal drift.
On the other hand, Fig.~\ref{Figure3}~(b) shows that for a gradient in 
$\gamma$, the anomalous drift is observed for the entire range of
$\Delta\gamma$ that is
investigated.

\begin{figure}
\centerline{
\includegraphics[width=8cm]{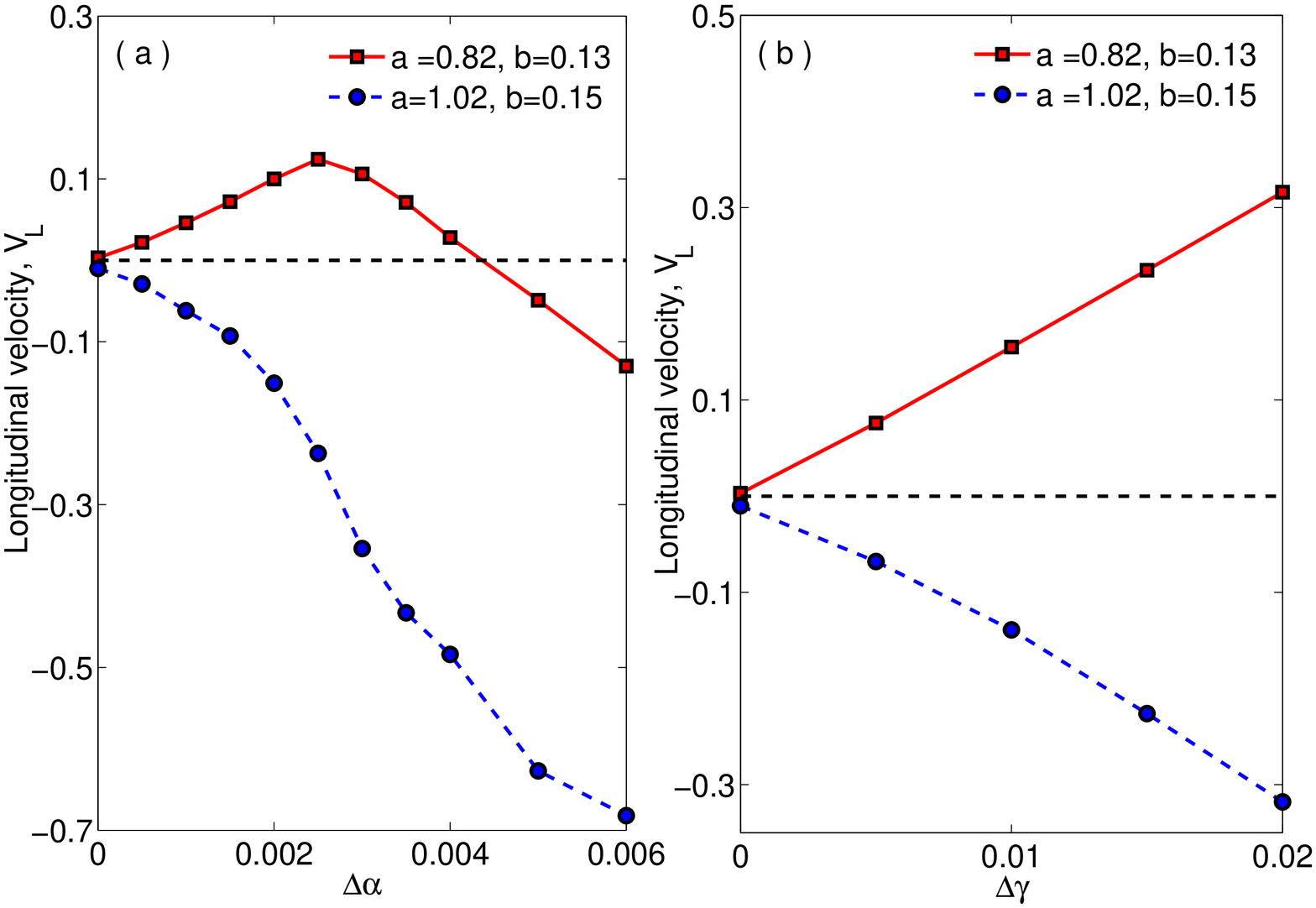}
}
\caption[]{
{\bf Drift velocity depends on the gradient in parameters $\alpha$ 
and $\gamma$.
}
(a) Non-monotonic variation (solid curve) of the longitudinal component 
of spiral
wave drift
velocity $V_L$ as a function of the gradient in local excitability,
$\Delta \alpha$, for a model system with parameters $a = 0.82, b =
0.13$.
Positive values of $V_L$ indicate anomalous drift. For a different set
of parameters ($a = 1.02, b = 0.15$),
normal drift is seen for the entire range of gradients used (broken curve).
(b) Variation of $V_L$ with the gradient in cellular coupling, 
$\Delta\gamma$. Solid and broken curves represent the anomalous
and normal drift seen for the two parameter sets mentioned earlier
(respectively), and are observed throughout the range of gradients used.
}
\label{Figure3}
\vspace{-0.6cm}
\end{figure}

We have also studied the effect of the local kinetics on anomalous drift 
by varying the Barkley model parameter $a$ (Fig.~\ref{Figure4},~a).
Increasing $a$ (keeping $b$ fixed) decreases the 
activation threshold of the medium, and thus makes the system more excitable. 
We observe that for both $\alpha$ and $\gamma$ gradients, the
variation of $V_L$ as a function of $a$ is non-monotonic. 
For the cellular coupling ($\gamma$) gradient,
the presence of anomalous regime clearly correlates with excitability. 
The drift is anomalous at lower excitability, but becomes normal 
at higher excitability. 
However, for the gradient in $\alpha$, the anomalous drift occurs only over 
an intermediate range of $a$. For lower and higher excitabilities, the
drift becomes normal.
\begin{figure}
\centerline{
\includegraphics[width=8cm]{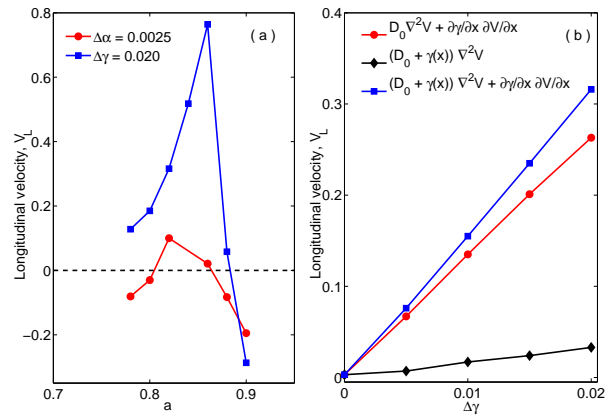}
}
\caption[]{
(a) Non-monotonic variation of the longitudinal component of 
 drift velocity, $V_L$, as a function of the model parameter $a$.
The two curves correspond to media having a constant gradient in excitability
(circles: $\Delta \alpha = 0.0025$, $\Delta \gamma = 0$) and cellular coupling 
(squares: $\Delta \alpha = 0$, $\Delta \gamma = 0.020$).
For both, the parameter $b = 0.13$.
(b) The contribution to $V_L$ from the different components in 
the diffusion term 
as a function of the gradient in cellular coupling, $\nabla \gamma$.
The circles and diamonds correspond to the linear and second-order 
contributions, and, squares correspond to the complete Laplacian term, 
respectively.
All data points shown are for $a = 0.82$, $b = 0.13$.}
\label{Figure4}
\vspace{-0.5cm}
\end{figure}

The mechanism of anomalous drift in $\alpha$ gradient remains unclear. 
However, we can understand the anomalous drift for the 
cellular coupling ($\gamma$) gradient, by relating it to other 
drift phenomena in excitable media.
Note that, the Laplacian term in Eq.~(\ref{eq1})
can be expanded as,
 \begin{equation}
        {\nabla}\gamma (x) D{\nabla}V = 
(D_0 + \gamma (x)) {\nabla}^2 V + \partial  \gamma / \partial x ~
\partial V/ \partial x.
\label{eq3}
  \end{equation}
Therefore, the heterogeneous cellular coupling $\gamma (x)$ 
contributes to both the gradient 
($\partial  \gamma / \partial x ~ \partial V/ \partial x$), as well as, 
second order
($\gamma (x) {\nabla}^2 V$) terms.
The relative contributions of these terms to the longitudinal
component of drift velocity is shown in Fig.~\ref{Figure4}~(b). 
We observe that the principal effect on $V_L$ is due to the
$\partial  \gamma / \partial x ~ \partial V/ \partial x$ term, while
$\gamma (x) {\nabla}^2 V$ accounts only for about $10 \%$ of the observed 
drift. This allows us
to propose the following explanation for anomalous drift in the presence of a
gradient in $\gamma$. 
If we do not consider the 
$\gamma (x) {\nabla}^2 V$ term in the Laplacian, the spatial operators 
in Eq.~(\ref{eq3}) are seen to
be identical to those in equations describing drift of a spiral wave 
in the presence of an electric field~\cite{Krinsky96}. This latter, in turn, 
is similar to the Laplacian describing the drift of radially symmetric 
filaments of a scroll ring in three-dimensional excitable 
media~\cite{Panfilov87, Alonso08}. As shown in Refs.~\cite{Henry02, Henry04},
the drifts observed in these two kinds of systems are induced by the
same instabilities.
We see from Fig.~\ref{Figure4}~(b) that the gradient 
$\partial  \gamma / \partial x ~ \partial V/ \partial x$, which determines 
the drift in an electric field and that of scroll wave filaments, 
also determines the drift as a result of $\gamma$ gradient. Therefore, 
we infer that the anomalous drift direction (towards higher excitability) 
observed by us is also a result of the same  long wavelength instabilities 
determining the drift of scroll wave filaments.
This suggests that the occurrence of scroll expansion in 3-D 
implies the existence of anomalous drift in $\gamma$ gradient in 2-D.
Conversely, observation of anomalous drift
might suggest parameter regions where scroll wave expansion is possible.

In this paper, we have explicitly demonstrated the possibility of 
spiral waves to drift towards region of higher excitability in a 
simple model of heterogeneous excitable medium. 
Most of the detailed ionic models for cardiac
tissue have the same form as Eq.~(\ref{eq1}) and, therefore, our analysis 
can be easily extended to biologically realistic models, such as LR1
or TNNP~\cite{Luo91,tenTuss04}. 
It might be possible to infer the parameter range in realistic models
where anomalous drift may occur by using the relation between the cellular
coupling gradient induced drift and scroll ring expansion. 
Note that, the latter
phenomenon has recently been seen 
in the LR1 model~\cite{Alonso08}.

Spiral waves are not only relevant for cardiac tissue,
but are also observed in many different excitable media. Thus, it may be possible
to relate our observation with results of kinematic studies~\cite{Mikhailov94} 
and models of cyclic catalysis in replicating entities~\cite{Pauline95},
which predict drift towards region with shorter periods.
From a clinical perspective, anomalous drift is important as it clearly
promotes arrhythmia and may result in fibrillation by promoting wave-breaks 
away from the spiral core.
Spiral drift in the presence of a cellular coupling gradient maybe 
a key factor giving rise to abnormal wave activity 
in regions of the heart where conductivity changes abruptly, e.g., 
at Purkinje-muscle cell junctions or in an infarct border
zone~\cite{Pumir05}. 
It can also be studied experimentally and numerically in many 
model systems, such as, heterogeneous mono-layers of neonatal rat 
cardiomyocytes~\cite{Biktashev09}. 
\vspace{-0.2cm}

To conclude, we have observed that spiral waves in heterogeneous excitable
media can drift towards regions having higher excitability. Such anomalous
drift occurs either in media having intermediate to low
excitability when the heterogeneity is a gradient in ionic current,
or, in media with low excitability for a gradient in the cellular
coupling. Further, it appears to be related to regimes
where expansion of 3-dimensional scroll wave filaments is observed.
Anomalous drift of spiral waves may increase the likelihood of the onset 
of complex spatio-temporal patterns in excitable medium, e.g.,
turbulent electrical activity in the heart.

This work was supported in part by Utrecht University 2008 Short-Stay 
Fellowship, IMSc Complex Systems Project (XI Plan) and IFCPAR Project
3404-4.

\bibliography{thebibliography,panfilov}

\begin{thebibliography}{10}

\bibitem{Keener98}
J. Keener and J. Sneyd, {\it Mathematical Physiology} 
(Springer, New York) (1998).

\bibitem{Davidenko91}
J. M Davidenko, {\it et al}, 
Nature {\bf 355}, 349 (1991). 

\bibitem{Sinha01} S. Sinha, A. Pande and R. Pandit, Phys. Rev. Lett. {\bf 86},
3678 (2001); S. Sridhar and S. Sinha, Europhys. Lett. {\bf 81},
50002 (2008).

\bibitem{Gray95}
R. A. Gray {\it et al}, Circulation {\bf 91}, 2454 (1995).

\bibitem{Barkley92}
D. Barkley, Phys. Rev. Lett. {\bf 68}, 2090 (1992).

\bibitem{Shajahan07}
T. K. Shajahan, S. Sinha and R. Pandit, Phys. Rev. E {\bf 75}, 011929
(2007).

\bibitem{PanfH:93}
A.~V. Panfilov and A.~V. Holden, J. Theor. Biol. {\bf 161},  271  (1993).

\bibitem{PanfKen:93c}
A.~V. Panfilov and J.~P. Keener, J. Theor. Biol. {\bf 163},  439  (1993).

\bibitem{Garfinkel2000}
A. Garfinkel and Z. Qu, in {\it Cardiac Eletrophysiology: From Cell 
to Bedside}, edited by D.~P. Zipes and J. Jalife
(Saunders, Philadelphia, 2004), p. 327.

\bibitem{Oxford05}
S. M. Cobbe and A. C. Rankin, in {\it Oxford Textbook of Medicine}, edited by
D. A. Warrell, T. M. Cox and J. D. Firth and E. J. Benz
(Oxford University Press, USA, 2005), p. 975.

\bibitem{Krinsky68}
V. I. Krinsky, Problemy Kibernetiki {\bf 2}, 59 (1968).

\bibitem{Pertsov90}
V. G. Fast and A. M. Pertsov, Biophysics {\bf 35}, 489 (1990).

\bibitem{Markus92}
M. Markus, Zh. Nagy-Ungavarai and B. Hess, Science {\bf 257}, 225 (1992).

\bibitem{Panfilov91}
A. V. Panfilov and B. N. Vasiev, Physica D {\bf 49}, 107 (1991).   

\bibitem{Panfilov83}
A. N. Rudenko and A. V. Panfilov, Stud. Biophysics {\bf 1}, 183 (1983).

\bibitem{Panfilov03}
K. H. W. J. ten Tusscher and A. V. Panfilov, Am. J. Physiol. 
Heart. Circ. Physiol {\bf 284}, H542 (2003).

\bibitem{Ivanitsky89}
G. R. Ivanitsky, V. I. Krinsky, A. V. Panfilov and M. A. Tsiganov, 
Biofizika {\bf 34}, 297 (1989).

\bibitem{Mikhailov94}
A. S. Mikhailov, V. A. Davydov and V. S. Zykov, Physica D {\bf 70},  1  (1994).

\bibitem{Gray96}
R. A. Gray, A. M. Pertsov and J. Jalife, Circulation {\bf 94}, 2649 (1996).

\bibitem{Jalife02}
J. Jalife {\it et al}, Cardiovasc. Res. {\bf 54}, 204 (2002).


\bibitem{Panfilov09}
R. H. Keldermann {\it et al}, Am. J. Physiol. Heart Circ. Physiol. 
{\bf 296},  H370  (2009).

\bibitem{Barkley90}
D. Barkley {\it et al}, Phys. Rev. A {\bf 42}, 2489 (1990).

\bibitem{Pumir09}
A. Pumir {\it et al}, arXiv:0902.3891

\bibitem{note1}
For $a,b$ parameters where anomalous drift is observed, the period and
wavelength of spiral wave exhibits a more rapid divergence with 
decreasing $\alpha$ compared to the parameter regime showing normal drift.

\bibitem{Krinsky96}
V. I. Krinsky, E. Hamm and V.Voignier, Phys. Rev. Lett {\bf 76},  3854  (1996).


\bibitem{Panfilov87}
A. V. Panfilov and A. N. Rudenko, Physica {\bf 28D},  215  (1987).

\bibitem{Alonso08}
S. A. Alonso and A. V. Panfilov, Phys. Rev. Lett {\bf 100}, 218101 (2008).

\bibitem{Henry02}
H. Henry and V. Hakim, Phys. Rev. E {\bf 65}, 046235 (2002).

\bibitem{Henry04}
H. Henry, Phys. Rev. E {\bf 70}, 026204 (2004).

\bibitem{Luo91} 
C. Luo and Y. Rudy, Circ. Res. {\bf 68}, 1501 (1991).

\bibitem{tenTuss04}
K. H. W. J Ten Tusscher, D. Noble, P. J. Noble and A. V. Panfilov, 
Am. J. Physiol. Heart. Circ. Physiol. 
{\bf 286}, H1573 (2004).

\bibitem{Pauline95}
M. C. Boerlijst and P. Hogeweg, Physica D {\bf 88}, 29 (1995).

\bibitem{Pumir05}
A. Pumir {\it et al}, Biophys. J. {\bf 89}, 2332 (2005).

\bibitem{Biktashev09}
V. N. Biktashev {\it et al}, Biophys. J.
{\bf 94}, 3726 {2009}.

\end{thebibliography}

\end{document}